\documentclass[prl,aps,showpacs,twocolumn]{revtex4}
\usepackage{amsmath}
\usepackage{graphicx}
\begin{document}
\title{Production of nonlocal quartets and phase-sensitive entanglement in a
  superconducting beam splitter}
\date{\today}

\author{Axel Freyn$^{1}$, Benoit Dou\c{c}ot$^{2}$, Denis Feinberg$^{1}$,
  R\'egis M\'elin$^{1}$}
\email{Regis.Melin@grenoble.cnrs.fr}
\affiliation{$^{1}$ Institut NEEL, CNRS and Universit\'e Joseph Fourier, BP 166,
  F-38042 Grenoble Cedex 9, France}
\affiliation{$^{2}$ Laboratoire de Physique Th\'eorique et des Hautes
  Energies, CNRS UMR 7589, Universit\'es Paris 6 et 7, 4 Place Jussieu, 75252
  Paris Cedex 05}

\begin{abstract}
Three BCS superconductors S$_a$, S$_b$, and S and two short normal regions
N$_a$ and N$_b$ in a three-terminal S$_a$N$_a$SN$_b$S$_b$ setup provide a 
source of nonlocal quartets spatially separated as two correlated pairs in
S$_a$ and S$_b$, if the distance between the interfaces N$_a$S and SN$_b$ is
comparable to the coherence length in S. Low-temperature dc transport of
nonlocal quartets from S to S$_a$ and S$_b$ can occur in equilibrium, and
also if S$_a$ and S$_b$ are biased at opposite voltages. At higher
temperatures, thermal excitations result in correlated current fluctuations
which depend on the superconducting phases  $\phi_a$ and $\phi_b$ in S$_a$ and
S$_b$. Phase-sensitive entanglement is obtained at zero temperature if N$_a$
and N$_b$ are replaced by discrete levels.
\end{abstract}
\maketitle

Regarding the manipulation of entangled states, quantum nanoelectronics is on
the way to address the same fundamental issues with electrons as quantum
optics does with photons. An entangled quantum state has a density matrix
distinct from that of any ``hidden-variable'' theory. Two-particle
entanglement can be probed~\cite{Aspect} via the violation of the Bell
inequality~\cite{Bell}. Multiparticle entanglement also has a high potential,
for instance it~\cite{Zhao} can be
used to implement error correction codes. 

Concerning superconductivity, two-particle entanglement can be generated at
 normal metal-superconductor NSN interfaces, by extracting a split Cooper pair 
from the BCS condensate of electron pairs~\cite{Byers-Flatte,Martin0,Deutscher-Feinberg,Choi}. We show in this Letter that a nanoscale
three-terminal superconducting setup can produce nonlocal quartets separated
as two pairs in different electrodes, therefore opening a route for a new
generation of entanglers which could be controlled by 
an
electromagnetic field. One must stress that here quartets are absent in the bulk 
superconductors which instead carry ordinary BCS pairing. This is in contrast with the destruction of the ``ordinary'' pair condensate 
in certain arrays of Josephson junctions~\cite{Doucot}, and its relationship with 
topological quantum computation~\cite{Ioffe,Kitaev}.  
Microscopically, nonlocal quartet transmission appears here as a generalization of
so-called Cooper pair splitting at a double NSN interface, and it can be
characterized by interference and noise. Recall that an Andreev pair in a
normal metal electrode N results from the emission of a charge $2e$ from S at
a NS interface, by the process of Andreev reflection
(AR). At a double NSN interface, spin-entangled~\cite{Lesovik2} or 
energy-entangled~\cite{Martin} pairs
can be produced through crossed (or nonlocal) Andreev reflection CAR~\cite{Manips,Manips-dots,Manips-noise}
 involving evanescent quasiparticle
states in S, on the coherence length $\xi_S$. CAR coexists with normal
transmission through S without electron-hole conversion (elastic cotunneling
EC)~\cite{Falci}. CAR or EC can be selected by their different Coulomb
interaction energy~\cite{Madrid-Delft}, by their spin
sensitivity~\cite{Deutscher-Feinberg,Falci}, by their distinguishing energy
dependence~\cite{Brinkman}, or by their different signature in the nonlocal conductance and in 
the zero-frequency shot
noise cross-correlations~\cite{Martin0, Lesovik2,Samuelsson-Buttiker,Bignon}. The new effects
considered here do not require more advanced technology than the experiments
on split pairs already realized with metallic structures~\cite{Manips} or with quantum
dots~\cite{Manips-dots}. 

\begin{figure}[tb]
\centerline{\includegraphics[width=0.7\columnwidth]{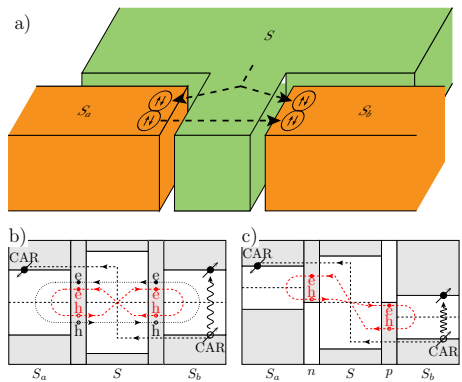}}
\caption{(Color online). Panel (a) shows a S$_a$SS$_b$ structure, where N$_a$
  and N$_b$ have not been represented for clarity. The processes
  taking place in three-terminal S$_a$N$_a$SN$_b$S$_b$ [panel (b)] and
  S$_a$$n$S$p$S$_b$ [panel (c)] structures are as follows: double crossed
  Andreev reflection (dCAR, red long dashed lines) producing a nonlocal
  quartet (a spatially separated pair of pairs), double elastic cotunneling
  (dEC, dotted black lines) exchanging pairs between S$_a$ and S$_b$, and
  crossed Andreev reflection (CAR), thermally activated above the gaps of
  S$_a$ and S$_b$. In addition, elastic cotunneling (EC) and local Andreev
  reflection (AR), not shown in the figure, also take place in
  S$_a$N$_a$SN$_b$S$_b$. With suitable gate voltages, the $n$- and $p$-doped
  semiconductors on panel (b) have a vanishingly small density of states at
  negative and positive energies, respectively. They filter CAR and dCAR, and
  eliminate EC, dEC, and AR.
  \label{fig:schemas}}
\end{figure}

In this Letter, a route to the production of nonlocal quartets is
proposed on the basis of bunching of two Andreev pairs in a superconducting
beam splitter made of conventional BCS superconductors.  Indeed, two Josephson 
junctions separated by a distance $d_S$ of the order of the coherence length $\xi_S$ of S
can be coupled by nonlocal coherent effects~\cite{Shafranjuk}. Here we study 
microscopically all the possible nonlocal effects and discuss their physical consequences. 
Remarkably, in a three-terminal S$_a$N$_a$SN$_b$S$_b$ structure,
nonlocal quartets can be separately transmitted as two correlated pairs in
S$_a$ and S$_b$. Non-local quartet transmission proceeds through double crossed
Andreev reflection (dCAR), which coexists with double elastic cotunneling dEC. The latter 
process yields Cooper pair transmission between S$_a$ and S$_b$~\cite{Freyn}. The
phase-sensitive dCAR is a new coherent nonlocal quantum channel, which has no
direct analog for incoherent multiple Andreev reflections~\cite{Houzet}. The
elementary charges involved in dCAR and dEC are doubled as compared to CAR and
EC. The four processes of CAR, EC, dCAR and dEC (see Fig.~\ref{fig:schemas})
will be treated on an equal footing~\cite{note-ordre}, as well as AR at each
of the S$_a$N$_a$S or SN$_b$S$_b$ interfaces which transfers Cooper pairs
between
S and S$_a$ or S$_b$. 

Cooper pair splitting in a S$_a$N$_a$SN$_b$S$_b$ structure with arbitrary
interface transparency can be described by generalizing the Andreev--Kulik--Saint-James bound
states (ABSs)~\cite{Andreev,Kulik,Saint James,Beenakker}. Those states are
coherent superpositions of electrons and holes, forming in a short single
channel junction a doublet at opposite energies $(E_-,E_+)$, and carrying
Josephson currents in opposite directions. In the S$_a$N$_a$SN$_b$S$_b$ setup, 
the phases of  S$_a$, S$_b$ and S being $\phi_a$, $\phi_b$ and
$\phi_S$, and taking the bias voltages V$_a$, V$_b$ to be zero, the total
currents $I_a$, $I_b$ are given at zero temperature $T=0$ by  $I_a(\phi_a, \phi_b) \approx
(2e/\hbar) \sum_{n=1,2}(\partial E_n/\partial\phi_a)$,  $I_b(\phi_a, \phi_b) \approx
(2e/\hbar) \sum_{n=1,2}(\partial E_n/\partial\phi_b)$. The $E_n$ are the energies of the 
ABSs formed by hybridizing the ABSs of both junctions through dCAR and dEC. 
They depend on both phase differences
$\delta \phi_a = \phi_a-\phi_S$ and $\delta \phi_b = \phi_b-\phi_S$. If $d_S
\alt \xi_S$, this leads (at lowest order in the dCAR and dEC processes) to
$I_a = I_a^0(\delta \phi_a) + I^{\mathrm{dEC}}(\delta \phi_a - \delta \phi_b)
+ I^{\mathrm{dCAR}}(\delta \phi_a + \delta \phi_b)$ and $I_b = I_b^0(\delta
\phi_b) - I^{\mathrm{dEC}}(\delta \phi_a-\delta \phi_b) +
I^{\mathrm{dCAR}}(\delta \phi_a + \delta \phi_b)$. Production of nonlocal
quartets (dCAR) and pair transmission (dEC) couple the coherent dc Josephson
currents in S$_a$ and S$_b$ by the inverse crossed {\it inductances}
$(L^{-1})_{a,b} = \partial I_a(\phi_a,\phi_b)/\partial\phi_b$ and
$(L^{-1})_{b,a} = \partial I_b(\phi_a,\phi_b)/\partial\phi_a$, which is an
extension of the concept of crossed {\it conductances} in a N$_a$SN$_b$
structure.
 
Voltages $V_a$, $V_b$ are applied now on the electrodes S$_a$, S$_b$ 
($V_S=0$ is the reference voltage). Yet a dc Josephson
current can flow from S to S$_a$ and S$_b$ if $d_S\sim\xi_S$~\cite{EPAPS},
in addition to the standard ac Josephson currents. The intensity of this current
can be seen as a synchronization of the phases of the
ac oscillations. Indeed, considering for simplicity low transparency contacts, 
the double Josephson junction is described by the
Hamiltonian ${\cal H} = {\cal H}_{\rm loc} + {\cal H}_{\mathrm{dCAR}} + {\cal
  H}_{\mathrm{dEC}} + 2e (\hat{n}_S V_S + \hat{n}_aV_a + \hat{n}_b V_b)$, with
${\cal H}_{\rm loc}=-E_J[\cos(\delta \phi_a(t))+\cos(\delta \phi_b(t))]$, and
${\cal H}_{\mathrm{dCAR}}=-E_J^{\mathrm{dCAR}}\cos(\delta \phi_a(t)+\delta
\phi_b(t))$, ${\cal H}_{\mathrm{dEC}}= - E_J^{\mathrm{dEC}}
\cos(\delta\phi_a(t) - \delta\phi_b(t))$. The currents are obtained from
Hamilton equations, for instance:
\begin{align}
&\left(\frac{\hbar}{2e}\right) I_a(t) = E_J\sin[\delta\phi_a(t)] +\\
\nonumber
&E_J^{\mathrm{dCAR}} \sin[\delta \phi_a(t) + \delta \phi_b(t)]
+ E_J^{\mathrm{dEC}} \sin[\delta \phi_a(t) - \delta \phi_b(t)].
\end{align}
Applying opposite voltages $V_a=-V_b\equiv V$ leads to a dc Josephson effect
for nonlocal quartets because the phase combination $\delta \phi_a(t) +
\delta\phi_b(t) = e(V_a+V_b)t/\hbar + \delta\phi_a + \delta\phi_b$ is then
time independent. The Josephson effect for nonlocal quartets becomes ac only
if the energy $eV_a+eV_b$ acquired by the quartet when separately transmitted
into S$_a$ and S$_b$ is finite. This result holds for any transparency.
  
The dc Josephson effect for nonlocal quartets is further considered for a
S$_a$$n$S$p$S$_b$ junction biased at opposite voltages, where the previous
N$_a$ and N$_b$ metals have been replaced by $n$- and $p$-doped
semiconductors. The conduction band edge on one side ($n$ type) and
the valence band edge on the other ($p$ type) are at zero energy. The gaps in
the density of states of the $n$- and $p$-doped semiconductors filter the
processes with positive energies in $n$, and with negative energies in
$p$~\cite{Brinkman} [Fig.~\ref{fig:schemas}(c)]. This excludes both the local Josephson effect and the
nonlocal dEC, thus leaving at $T=0$ only the nonlocal dCAR as a coherent
coupling between the condensates~\cite{note-CAREC}. In this ideal situation,
one obtains a perfect superconducting beam splitter operating at the scale of
the coherence length, and producing correlated pairs of Cooper pairs flowing
in the leads S$_a$ and S$_b$. Notice again that the biases V$_a$ and V$_b$
should be {\it opposite} in the coherent Josephson regime, while they are equal in a
normal beam splitter N$_a$NN$_b$ or a Cooper pair splitter N$_a$SN$_b$, where
quasiparticles are emitted instead of pairs. 

Let us now discuss the noise cross-correlations in a S$_a$N$_a$SN$_b$S$_b$ 
structure. Zero-frequency thermal noise is present in the
absence of applied voltage for sufficiently transparent single
junctions~\cite{Martin-Rodero,Imam}. Finite values are obtained for all the
components of the correlators $S_{i,j}(t) = \langle \delta I_i(t+t')\delta
I_j(t)\rangle$, where $\delta I_k$ is the current fluctuation in S$_k$
($k=a,b$). This equilibrium noise, due to thermally activated fluctuations
between ABSs carrying opposite currents, is phase sensitive because the
population of quasiparticles in thermal equilibrium exchanges charge with the
condensate. The thermal noise can be very large for a long inelastic lifetime
of the ABSs~\cite{Martin-Rodero,Imam}. CAR and EC lead to components
$S_{a,b}^{\mathrm{CAR}}$ and $S^{\mathrm{EC}}_{a,b}$ of $S_{a,b}$, which are
independent of $\delta \phi_a$ and $\delta \phi_b$. They correspond to CAR and
EC assisted by thermal activation over the gap $\Delta$
(Fig.~\ref{fig:schemas}). On the contrary, dCAR and dEC result in thermal
fluctuations between hybridized ABSs. dCAR corresponds to random emission and
absorption of nonlocal quartets between S and S$_a$, S$_b$. dEC corresponds
to random transmission of pairs between S$_a$ and S$_b$. The contributions
$S^{\mathrm{dCAR}}_{a,b}$ and $S^{\mathrm{dEC}}_{a,b}$ of dCAR and dEC to
$S_{a,b}$ depend, respectively on the phase combinations $\delta \phi_a +
\delta \phi_b$ and $\delta \phi_a - \delta \phi_b$. Generalizing
Refs.~\cite{Imam,Martin-Rodero,Buttiker}, the zero-frequency noise
cross-correlations $S_{a,b}$ in the absence of applied voltage is written as
\begin{equation}
  S_{a,b}=\frac{e^2}{\eta \hbar^2}\sum_n \frac{1}{\cosh(E_{
   n}/2k_BT)}\frac{\partial E_{n}}{\partial \phi_a}\frac{\partial E_{n}}{\partial \phi_b}
,
\end{equation}
where the lifetime of the Andreev states $1/\eta$ shows the relevance of
pair-breaking effects in the superconductor~\cite{Shafranjuk}. Such noise
correlations could be large at temperatures $T^*$ comparable to the ABS energy
level difference. The crossover temperature $T^*$ is strongly reduced as the
interface transparency increases.

\begin{figure}[tb]
\centerline{\includegraphics[width=0.7\columnwidth]{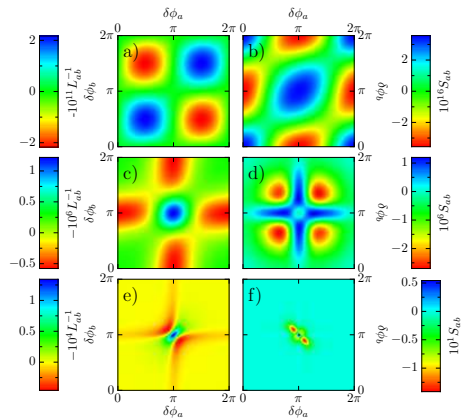}}
\caption{(Color online). The plots show $-(L^{-1})_{a,b}$ [panels (a), (c),
  (e)] and $S_{a,b}$ [panels (b), (d), (f)] as a function of
  $(\delta\phi_a,\delta\phi_b)$ for $T/\Delta=0.1$, $T_N=4.10^{-4}$ [panels
  (a),(b)], $T_N=0.64$ (panels c, d) and $T_N=1$ [panels (e), (f)]. The ratio
  $\Delta/\Delta_S=0.1$ is used.
\label{fig:plot}
}
\end{figure}

Calculations of the equilibrium nonlocal inverse inductance
$(L^{-1})_{a,b}(\delta \phi_a,\delta \phi_b)$ and cross-correlations
$S_{a,b}(\delta \phi_a,\delta \phi_b)$ are based on microscopic Nambu-Keldysh
Green's functions~\cite{EPAPS} in which the interfacial hopping
amplitude is accounted for by a self-energy. Arbitrary values of the
temperature and of the normal-state transmission coefficient $T_N$ can be
treated in equilibrium because the time convolutions in the Dyson equations
then simplify into products of Green's function depending only on energy. 
The microscopic calculations carried out for a three-dimensional
ballistic superconductor apply to a voltage range in which the proximity
effect is negligible. $S_{a,b}$ goes to zero at $T=0$ because
it is thermally activated over the ABS gap. $(L^{-1})_{a,b}$ saturates at
low $T$ to the zero-temperature response of the condensate. As seen from
perturbation theory in the tunnel amplitudes, both dCAR and dEC contribute with
a positive value to $S_{a,b}(\delta\phi_a=0, \delta\phi_b=0)$. Thus, for
tunnel contacts, $S_{a,b}(\delta\phi_a=0, \delta\phi_b=0)$ is positive at the
small $T/\Delta=0.1$ [see Fig.~\ref{fig:plot}(b)]. The oscillations in
$-(L^{-1})_{a,b}(\delta \phi_a, \delta \phi_b)$ [Fig.~\ref{fig:plot}(a)] match
those of $S_{a,b}(\delta \phi_a, \delta \phi_b)$ (see
Fig.~\ref{fig:plot}b). They reflect the distinguishing phase dependences of
dCAR and dEC. Increasing $T_N$ has the effect of favoring the transmission of
pairs by dEC from S$_a$ to S$_b$ (or from S$_b$ to S$_a$), and disfavoring
their transmission as a pair of holelike quasiparticles by dCAR. The
cross-shaped variations of $(L^{-1})_{a,b}(\delta \phi_a,\delta \phi_b)$ and
$S_{a,b}(\delta\phi_a,\delta\phi_b)$ for intermediate $T_N$ can be understood
from the bound state energy level minima in the $(\delta\phi_a, \delta\phi_b)$
plane for $\delta\phi_a,\delta\phi_b\in\{0,\pi,2\pi\}$. 

This superconducting beam splitter also generates entanglement 
between pair numbers in the two branches, as pairs are emitted two by two. 
Indeed a split Josephson current
due to nonlocal quartets connects coherently pair number states 
$|N_a\rangle |N_S\rangle
|N_b\rangle$  with states $|N_a + 1\rangle |N_S - 2\rangle |N_b + 1\rangle$, 
entangling the added pairs in $S_a$ and $S_b$~\cite{EPAPS}.
To illustrate this, let us replace the N junctions by quantum dots D$_a$,
D$_b$. Indeed phase-sensitive entanglement is obtained in a
S$_a$D$_a$SD$_b$S$_b$ structure biased at voltages $V_a=-V_b\equiv V$ larger
than $\Delta$, with $\Delta_S \gg \Delta,\,eV$. In a first step, D$_a$ and
D$_b$ are supposed to carry spin-degenerate orbitals. In addition, the gate
voltages are such that D$_a$ (D$_b$) has a level at $eV_a$ ($eV_b$) but no
level at $-eV_a$ ($-eV_b$). Quasiparticles are transmitted from S$_a$ to
D$_a$ (from S$_b$ to D$_b$) but local Andreev processes between D$_a$ and S
(D$_b$ and S) are not possible. In the limit of large gaps, 
three terms contribute to the effective Hamiltonian of
the two coherently coupled levels at energies $eV_a$ (in D$_a$) and $eV_b$ (in
D$_b$):  (i) {\it The exchange of pairs between S$_a$ and D$_a$ (S$_b$ and
  D$_b$):} ${\cal H}_{{\mathrm{AR}},a(b)} = -\alpha [\exp(i\phi_{a(b)})
  c_{a(b),\uparrow}^+ c_{a(b),\downarrow}^+ + h.c.]$. (ii) {\it Cooper pair
splitting:} ${\cal H}_{\mathrm{CAR}} = -\beta [c_{a,\uparrow}^+
  c_{b,\downarrow}^+ + c_{b,\uparrow}^+ c_{a,\downarrow}^++h.c.]$. (iii) {\it
  Production of nonlocal quartets:} ${\cal H}_{\mathrm{dCAR}} = -\gamma
[c_{a,\uparrow}^+ c_{a,\downarrow}^+ c_{b,\uparrow}^+
  c_{b,\downarrow}^++h.c.]$. Exact diagonalizations of ${\cal
  H}_{{\mathrm{AR}},a} + {\cal H}_{{\mathrm{AR}},b} + {\cal H}_{\mathrm{CAR}}
+ {\cal H}_{\mathrm{dCAR}}$ result in an entangled ground state characterized
by a positive concurrence, which depends on the values of $\delta\phi_a$ and
$\delta\phi_b$ via the combination $\delta\phi_a+\delta\phi_b$ typical of
dCAR~\cite{EPAPS}.  The Coulomb interaction Hamiltonian is ${\cal H}_U =
U\hat{n}_{a(b),\uparrow} \hat{n}_{a(b),\downarrow}$, with
$\hat{n}_{a(b),\sigma}$ the number of spin-$\sigma$ electrons in dot a (in dot
b). As $U$ increases,  the zero-temperature concurrence of the ground state
remains finite because of virtual excitations coupling to dCAR.

Entanglement can also be more directly assessed by showing that no classical 
correlation can account for the crossed noise of nonlocal quartets. Let us
consider S$_a$D$_a$SD$_b$S$_b$ or S$_a$$p$S$n$S$_b$ setups, biased at
opposite $V_a$ and $V_b$ larger than $\Delta$, and with $\Delta_S \gg
\Delta,\,eV$ [assumption (A1)]. The notation $\langle N_a(t,\tau) \rangle_\rho$
stands for the average number of electrons transmitted into electrode $S_a$ in
the time interval $[t, t+\tau]$.
The average over the
hidden-variable density matrix $\rho = \int d\lambda
\rho_a(\lambda)\otimes\rho_b(\lambda)$ is noted $\langle \cdots \rangle_\rho$~\cite{Lesovik2}. We
make the assumption (A2) that each subsystem a and b is separately
described by quantum solid-state physics: only communication through S
corresponds to a hidden variable. The assumption A2 leads to $\langle \delta
N_a(t,\tau) \delta N_b(t,\tau)\rangle_{qu} =\langle \delta N_a(t,\tau) \delta
N_b(t,\tau)\rangle_\rho $, with
\begin{eqnarray}
\langle \delta N_a(t,\tau) &&\mbox{\hspace*{-.7cm}}\delta
N_b(t,\tau)\rangle_\rho\\
&=&\int d\lambda f(\lambda) \langle \delta N_a(t,\tau) \rangle_\lambda
\langle \delta N_b(t,\tau)\rangle_\lambda
\nonumber
,
\end{eqnarray}
where $f(\lambda)$ is the probability density of the hidden variable
$\lambda$. The value of $\delta N_a(t,\tau)$ for the specific value $\lambda$
of the hidden variable is denoted by $\langle \delta N_a(t,\tau)
\rangle_\lambda$.

An additional assumption (A3) concerning the setup is made in order to simplify
the discussion: the gap $\Delta_S$ is larger than the bandwidth $W$ of
the superconductors S$_a$ and S$_b$, and the linewidth broadening in S is
$\eta_S=0$. The three assumptions (A1), (A2), and (A3) imply that charge transport is
blocked at any temperature because the D$_a$ and D$_b$, or the $n$ and $p$
energy filters [assumption (A1)], suppress all Josephson processes taking place
locally within each subsystem a or b for any realization of the hidden
variable $\lambda$ [assumption (A2)]. No quasiparticle is transmitted from S to
S$_a$ (from S to S$_b$) within subsystem a (b) for any value of $\lambda$
(assumption A3). The equality $\langle \delta N_a(t,\tau) \delta
N_b(t,\tau)\rangle_{qu}=0$ is then obtained because $\langle \delta
N_a(t,\tau) \rangle_\lambda = \langle \delta N_b(t,\tau)
\rangle_\lambda=0$. The following two statements are in conflict: (i) The
cross-correlations are finite; and (ii) The nonlocal processes is a classical
communication (related to some hidden variables) rather than the quantum
mechanical CAR, dCAR.

For the S$_a$$n$S$p$S$_b$ structure considered above, the cross-correlations
are vanishingly small at zero temperature and the Bell-like argument does not
imply entanglement in this case. Indeed, the ground state of a Josephson
junction is entangled only if the total number of pairs is fixed~\cite{EPAPS}.  
However, the denomination {\it phase-sensitive entanglement} is
appropriate at zero temperature for the S$_a$D$_a$SD$_b$S$_b$ structure
because the cross-correlations are finite at $T=0$, and the Bell-like argument
is in agreement with the direct calculation of the concurrence, being also
finite.

To conclude, we propose to generate nonlocal quartets in the
solid state, by producing correlated pairs in a superconducting beam splitter
involving three superconductors. The nonlocal inductance and the
phase-sensitive thermal cross-correlations may be probed in future
experiments. Of particular interest is the possibility to obtain entanglement
at zero temperature from current-current cross-correlations in a
S$_a$D$_a$SD$_b$S$_b$ structure. Non-local Shapiro steplike experiments are
also promising for investigating dCAR and dEC, in the sense that resonances
due to dCAR or dEC could be obtained if the ac voltage oscillations are
synchronized in two incoming channels S$_a$ and S$_b$.


\begin{thebibliography}{99}
\bibitem{note-ordre} In perturbation theory in the normal-state transmission
  coefficient $T_N$, the production of nonlocal quartets (the synchronization
  of two Josephson currents of order $T_N$), appears at the order
  $(T_N)^2$. EC and CAR are also of order $(T_N)^2$.
\bibitem{note-CAREC} The thermally activated CAR and EC are ineffective at
  $T=0$ in S$_a$N$_a$SN$_b$S$_b$ and in S$_a$$n$S$p$S$_b$.
\end{thebibliography}
\end{document}